\documentclass{emulateapj} \usepackage{times}
\newcommand{\etal}{{et al.}}
\newcommand{\eg}{{\it e.g.,}}

\shorttitle{Morphologies in a Cluster at z=1.34} 
\shortauthors{Fu, Stockton \& Liu} 
\journalinfo{accepted for publication in ApJ}
\submitted{accepted for publication in ApJ}

\begin{document}

\title{Morphologies in a Cluster of Extremely Red Galaxies with Old Stellar 
Populations at z\,=\,1.34\footnotemark[1]}

\footnotetext[1]{Based in part on data collected 
at Subaru Telescope, which is operated by the National Astronomical Observatory of Japan. 
Some of the data presented herein were obtained at the W.M. Keck Observatory, which is 
operated as a scientific partnership among the California Institute of Technology, the 
University of California and the National Aeronautics and Space Administration. The 
Observatory was made possible by the financial support of the W.M. Keck 
Foundation.}

\author{Hai Fu, Alan Stockton, and Michael Liu\footnotemark[2]}
\affil{Institute for Astronomy, University of Hawaii, 2680 Woodlawn
 Drive, Honolulu, HI 96822}
 
\footnotetext[2]{Alfred P. Sloan Research Fellow}

\begin{abstract}
We have identified a clustering of red galaxies from deep optical/IR images obtained
as part of the Institute for Astronomy Deep Survey. Photometric  spectral-energy
distributions indicate that most of these galaxies comprise nearly pure old stellar 
populations at a redshift near 1.4, and Keck spectroscopy of the three brightest red
galaxies confirm this interpretation and give redshifts ranging from 1.335 to 1.338.
Four of the galaxies are close together on the sky and less than 30\arcsec\ from 
an $R=13.5$ star, and we have
obtained deep adaptive-optics imaging of this group. Detailed analysis and modeling
of the surface-brightness profiles of these galaxies shows that two are normal
ellipticals, one is an S0, and one appears to be an essentially pure disk of old stars, 
with no significant bulge. All four are highly relaxed, symmetric systems.  While
the old, bulgeless disk galaxy represents a type that is rare at the present epoch, the
other three galaxies have structural parameters that are essentially
indistinguishable from those
of present-day galaxies and differ only in the age of their stellar populations.
\end{abstract}


\section{Introduction}

Clusters of galaxies at high redshifts are important both because their number density as
functions of mass and redshift can constrain cosmological models and because they 
provide samples of galaxies in dense environments for studies of galaxy formation and 
evolution. X-ray surveys published to date, spectacularly successful at locating massive clusters
out to $z \sim 0.7$ (\eg\ \citealt{ebe01,mul03}), have not had the sensitivity to produce significant 
samples at $z>1$.  Surveys currently under way with XMM-Newton show promise of extending the
range for well characterized samples to $z\sim1.5$ (\eg\ \citealt{sch04,mul05}). In the meantime,
optical/IR approaches to cluster detection are undergoing a renaissance through the use of
photometric redshift information in multi-band large-area surveys to weed out contaminating 
foreground and background sources.

Of the definite clusters identified so far at $z>1$, five have been found in deep X-ray images
\citep{ros99,sta02,ros04,has04,mul05}. The majority of the others have been identified in
pointed observations of radio-source fields (\eg\ \citealt{hal98,liu00,bes03,tof03,wol03,ste03}). 
Although it has been well established that extremely red objects (EROs) show strong clustering
(\eg\ \citealt{dad03,roc03,geo05}), the only confirmed cluster so far reported at $z>1$
from general field surveys is ClG J0848+4453, at $z=1.27$ \citep{sta97}.
\citet{saw05} report a strong density enhancement of extremely red objects (EROs), comparable
to those seen in some radio-source fields, but they have only $R$ and $K_S$ imaging and
therefore can neither estimate a redshift nor even confirm a common redshift for their objects. 
A near-miss (as far as redshift is concerned) is the identification from optical/IR imaging 
(confirmed by spectroscopy) of a rich cluster at $z=0.95$ by \citet{bar04}.

We report here the discovery of an apparent cluster of EROs at $z=1.34$ from a field optical/IR survey.
Since four of the galaxies lie within 30\arcsec\ of a 13th mag star and very
close to one another, we have been able
to obtain adaptive-optics (AO) imaging to determine their detailed morphologies.
We assume a standard concordance cosmological model throughout, with
$H_0 = 70$ km s$^{-1}$, $\Omega_m = 0.3$, and $\Omega_{\Lambda}=0.7$.

\section{Observations, Data Reduction, and Results}

\subsection{Optical and IR Imaging}
The initial discovery of the cluster of red objects came from deep $R$, $I$, and $Z$ imaging 
obtained as part of the Institute for Astronomy Deep Survey (IfA-DS; we will refer to the cluster
as ``IDSCL\,J0749+1018'').  The IfA-DS was the
result of a proposal from a consortium of astronomers to obtain imaging in 5 fields totalling
2.5 square degrees.  The survey data were designed to be used for a wide variety of
research projects, including a supernova search \citep{bar04b}, galactic structure studies, a search
for brown dwarfs \citep{liu02}, and a search for EROs. The observations were obtained with the
SuprimeCam CCD mosaic camera, located at the prime focus of the 8 m Subaru
telescope. SuprimeCam comprises ten $2048\times4096$ CCDs and has a
$34\arcmin\times27\arcmin$ field of view. Each of the 5 survey fields was covered by
two slightly overlapping SuprimeCam fields, giving a final image of approximately
$34\arcmin\times53\arcmin$.  The supernova search constrained the design of the
survey, requiring observations spaced at a variety of time intervals. Details of the
survey, as well as the reduction procedures applied to produce the
final images, can be found in \citet{bar04b}.

Because of the vagaries of the weather and of seeing conditions, the final depth in each
of the filters varied for the different fields. For our ERO searches, we have concentrated 
on the two fields for which the deepest imaging was obtained, centered at 
$04^{\rm h} 38^{\rm m} 40^{\rm s}$, $-01\arcdeg\ 30\arcmin\ 00\arcsec$ (``0438 field'') and
$07^{\rm h} 49^{\rm m} 55^{\rm s}$, $+10\arcdeg\ 09\arcmin\ 00\arcsec$ (``0749 field'').
The limiting magnitudes for $R$, $I$, and $Z$ (Vega system, $5\sigma$, 1\arcsec-radius aperture) 
are, respectively, 26.8, 26.1, and 25.7 for the 0438 field and 26.6, 26.6, and 25.8 for 
the 0749 field. The general survey of these fields will be discussed elsewhere.

Sources in the 0749 field were identified by running SExtractor \citep{ber96} on the $Z$-band image,
using a Gaussian filter matched to the seeing profile. This procedure resulted in
a catalog of positions and magnitudes for 145058 sources in the 0.5 square degree field. Colors were found
by re-running SExtractor in the two-image mode, in which objects found in the $Z$-band image
were measured with the same aperture in the registered $R$ or $I$-band image. Catalogs
could then be extracted for various magnitude and color cuts.

Because of the limited fields of IR detectors available to us at the time, and because we wished
to develop a sample suitable for AO imaging studies of morphology, we focused initially either on 
fields with high densities of red objects or on red objects close to stars with $R\le14$.  
IDSCL J0749+1018 satisfies both criteria. We observed this and other fields of
interest with the CISCO IR camera on the Subaru telescope, obtaining imaging at $K'$
and $J$ to distinguish spectral-energy distributions (SEDs) of
galaxies comprising old stars from those of heavily reddened starbursts, the other main
contributor to the ERO population.  The observations were obtained on 2002 Feb 5 UT, 
with a total of 24 min integration on the cluster field at $K'$ and 16 min at $J$.
The seeing FWHM for the $K'$ image was 0\farcs84, and that for the $J$ image was
1\farcs1. The photometric calibration came from observations of the UKIRT faint standards
FS12, FS15, FS122, and FS125 \citep{haw01}. The cluster field is shown in Fig.~\ref{rik}, 
and an $I\!-\!K'$ vs. $K'$ color-magnitude diagram based on magnitudes determined 
with SExtractor is shown in Fig.~\ref{cmdiag}. Objects in the shaded band are potential
candidates for pure old stellar populations with effective ages between 1 Gyr and 3.5 Gyr.
We can refine this sample both by bringing to bear photometry in additional bandpasses
(which can, e.g., distinguish between the two general types of ERO mentioned above) and
by carefully optimizing the photometry. The points shown in red in Fig.~\ref{cmdiag}
are those for which this more extensive and more accurate dataset is consistent with
a spectral-energy distribution (SED) due to old stars and for which the final photometry indicates
a total $K'$ mag of $<20$ and $R\!-\!K'\gtrsim6$.

\begin{figure}[!t]
\epsscale{0.8}
\plotone{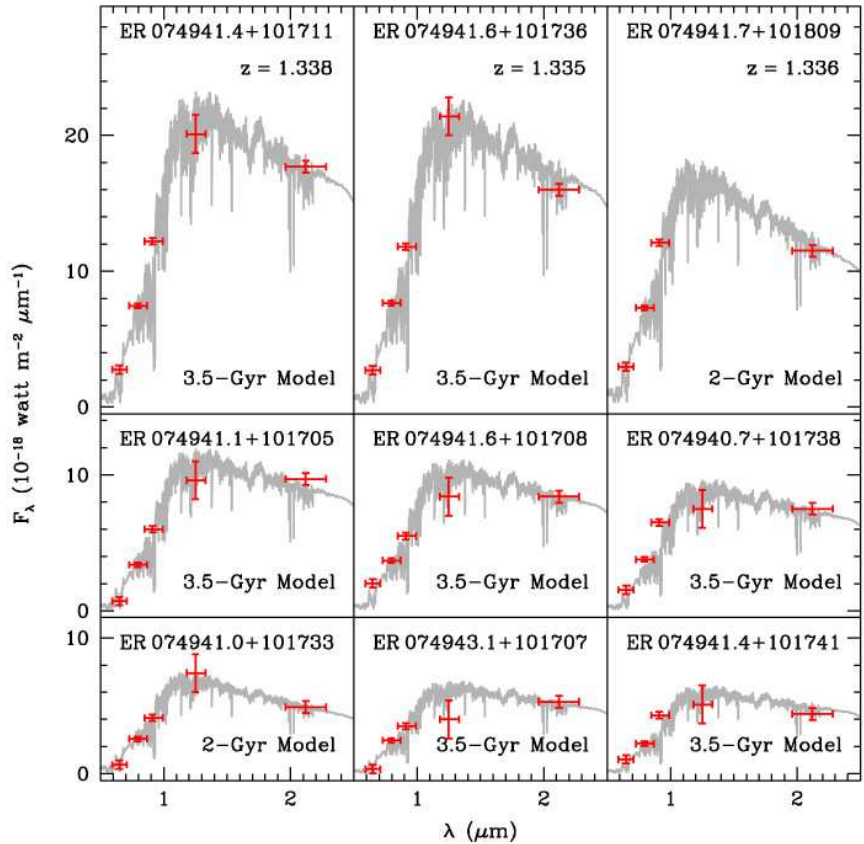}
\caption{Spectral-energy distributions of 9 of the reddest galaxies in the field
from optical/IR photometry.  The gray traces are reference Bruzual-Charlot (2003)
instantaneous-burst models with ages as given in each panel and redshifted to
$z=1.34$.  Horizontal bars
on the photometric points show filter half-transmission widths; vertical bars give
1-$\sigma$ random errors of the photometry. All of these galaxies, with the
exception of ER\,074943.1+101707 (which may be a dusty starburst), clearly 
show the strong inflection near
rest-frame 4000 \AA\ characteristic of old stellar populations.}\label{sed}
\end{figure}

The SEDs for these 9 galaxies with  are shown in
Fig.~\ref{sed}. \citet{bru03} solar-metallicity instantaneous-burst models with ages of 
3.5 or 2 Gyr are shown for comparison. No great weight should be given to the actual
ages implied because of the age-metallicity degeneracy and other issues; however, the 
generally good fits do indicate that most of these galaxies comprise truly old stellar populations
with little reddening or admixture of even a small amount of recent star formation.

It is clear from Fig.~\ref{rik} that there are additional red objects in the field;
many of these have $R\!-\!K'\gtrsim5$, which is sometimes taken as a definition for an ERO.
Here, however, we will retain the more stringent definition ($R\!-\!K'\ge6$, within the photometric 
errors) in order to define a cleaner sample of old stellar populations.

\subsection{Spectroscopy}
\begin{figure}[!tb]
\epsscale{0.8}
\plotone{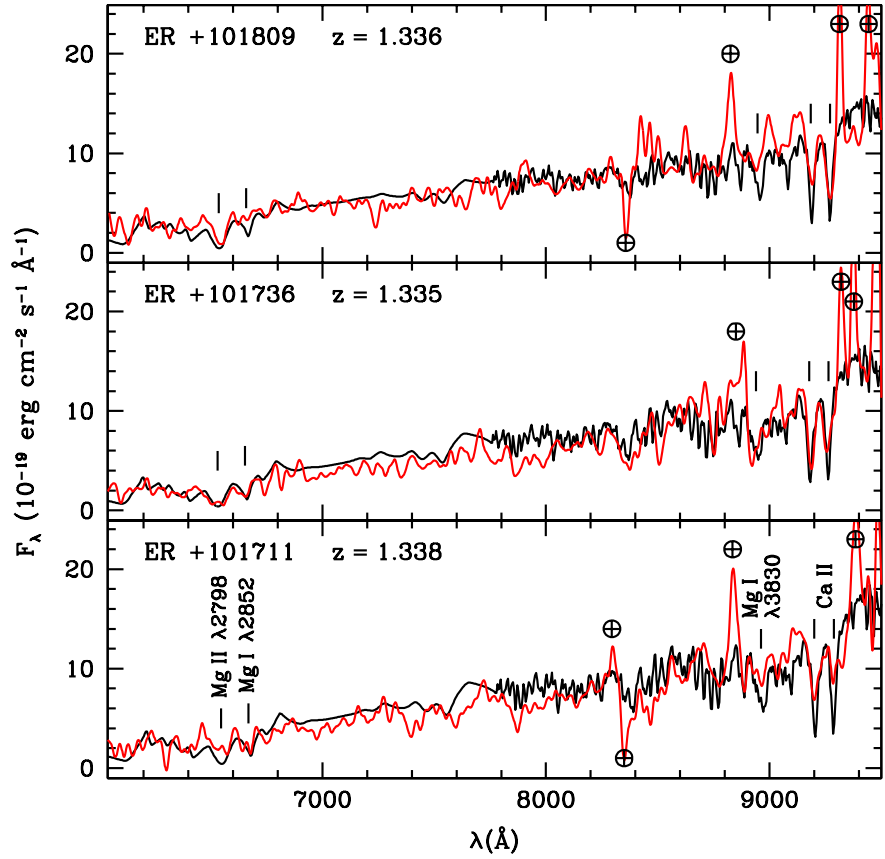}
\caption{Spectra of three galaxies in the cluster field (red lines), overlaid on 3.5 Gyr 
(bottom two panels) or 2 Gyr (top panel)
instantaneous-burst \citet{bru03} models (black lines).  Some spectral features in the galaxies 
are indicated by the vertical lines. The ``$\oplus$'' symbols indicate positions where
strong airglow lines, together with fringing in the CCD, have resulted in significant
sky-subtraction residuals. See text for details.}\label{spect}
\end{figure}

We obtained spectroscopy of the 3 brightest EROs in the cluster field with the
Low-Resolution Imaging Spectrograph (LRIS; \citealt{oke95}) on the Keck I telescope
on 2003 Jan 30 UT.
We used a multi-slit mask with slit widths of 1\farcs2. The typical seeing PSF during the
observations was 0\farcs65.
The 400 line mm$^{-1}$, 8500 \AA\ blazed grating gave a nominal resolution of about
10 \AA, and the total integration time was 250 min. The spectrophotometric calibration
was obtained from observations of the standards G191B2B and Feige 67.
The three galaxies all show spectra characteristic of old stellar populations and
have closely similar redshifts near 1.34 (Fig.~\ref{spect}).
The redshift fortunately places the strong H and K lines of \ion{Ca}{2} in the atmospheric 
window clear of strong airglow lines between 9000 \AA\ and 9300 \AA, leading to quite
firm redshifts in spite of occasionally large residuals from strong airglow lines elsewhere
in the spectra. These large residuals occur, mainly at the red end of the spectrum, because
the fringing pattern in the detector is time variable because of flexure in the instrument, and
it cannot entirely be removed.

The photometric and spectroscopic data for the galaxies are summarized in Table \ref{galprop}.
While the galaxy designations in the table give sufficiently accurate coordinates to uniquely 
identify the objects, we often use only the declination portion of the name in figures to avoid 
crowding.

\subsection{Adaptive-Optics Imaging}
\begin{figure}[!t]
\epsscale{1.0}
\plotone{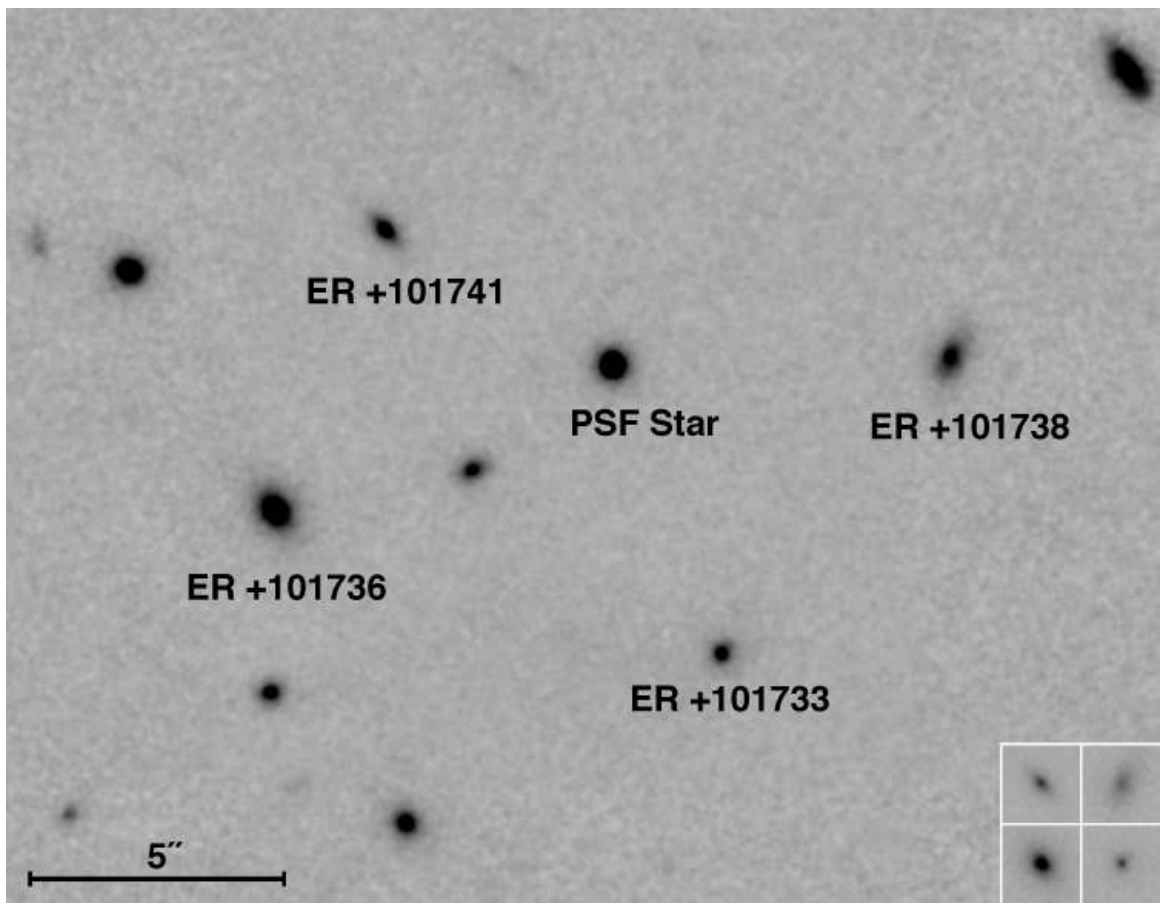}
\caption{The AO image obtained in the $K'$ band with the Subaru AO system and IRCS.
The ERO galaxies are labeled. The insets in the lower-right corner show lower-contrast
images in the same rotational order as the main frame. The 5\arcsec\ scalebar corresponds
to 42 kpc at $z=1.34$.  The main image has been 
smoothed with a Gaussian with $\sigma=2$ pixels; the
insets have been similarly smoothed with $\sigma=1$ pixel.}\label{ao}
\end{figure}

Finally, in order to determine morphologies of some of the galaxies, we carried out AO 
imaging in the $K'$ band of a portion of the cluster with the curvature-sensing
AO system on the Subaru telescope and the Infrared Camera and Spectrograph (IRCS;
\citealt{kob00}) on 2003 Feb 17 UT, using the 22.6 mas pixel$^{-1}$ scale. The intrinsic seeing
at $K'$ was 0\farcs6.  The total integration time was 135 min. The AO guide star
(USNO B1 1002-0150882) has an $R$ mag of 13.5, and the galaxies were at
distances ranging from 18\arcsec\ to 28\arcsec\ from the guide star.
The final image has a FWHM of 91 mas and is shown in Fig.~\ref{ao}. 

We used C. Y. Peng's GALFIT procedure \citep{pen02} to fit the two-dimensional  profiles of the
galaxies observed with the Subaru AO system. There is a star suitable for determining the PSF 
conveniently placed in the middle of the group of 4 extremely red galaxies that fit within our 
AO imaging field, and we find no discernable anisoplanatic distortion across the portion of the 
field including the galaxies.

\section{Surface-Brightness Profiles and Morphologies}

\subsection{ER\,074941.6+101736}
\begin{figure}[!t]
\epsscale{0.9}
\plottwo{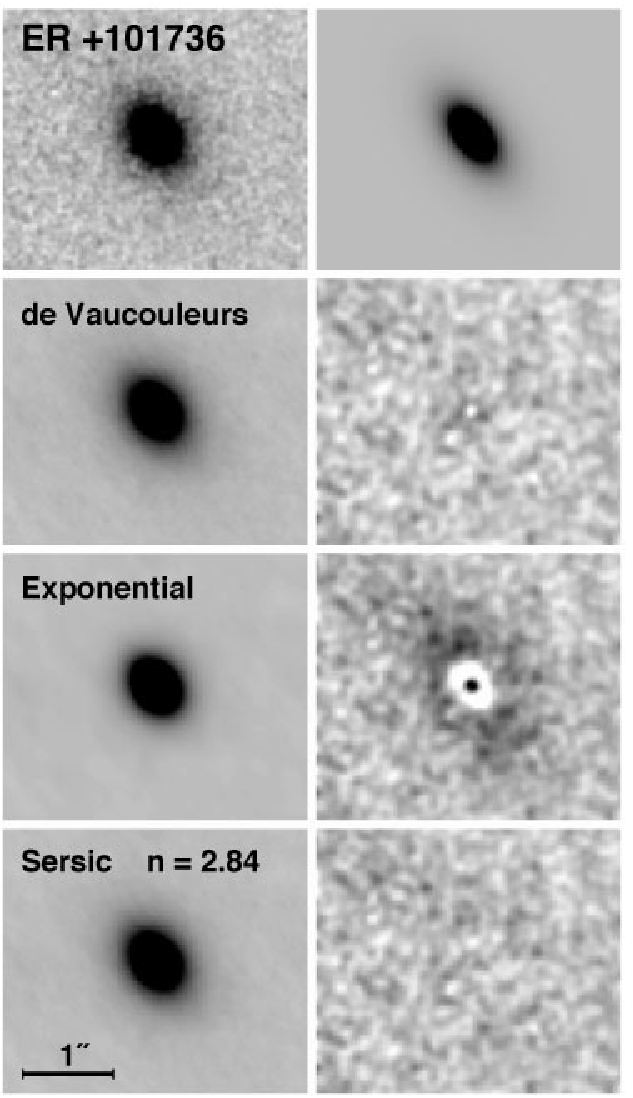}{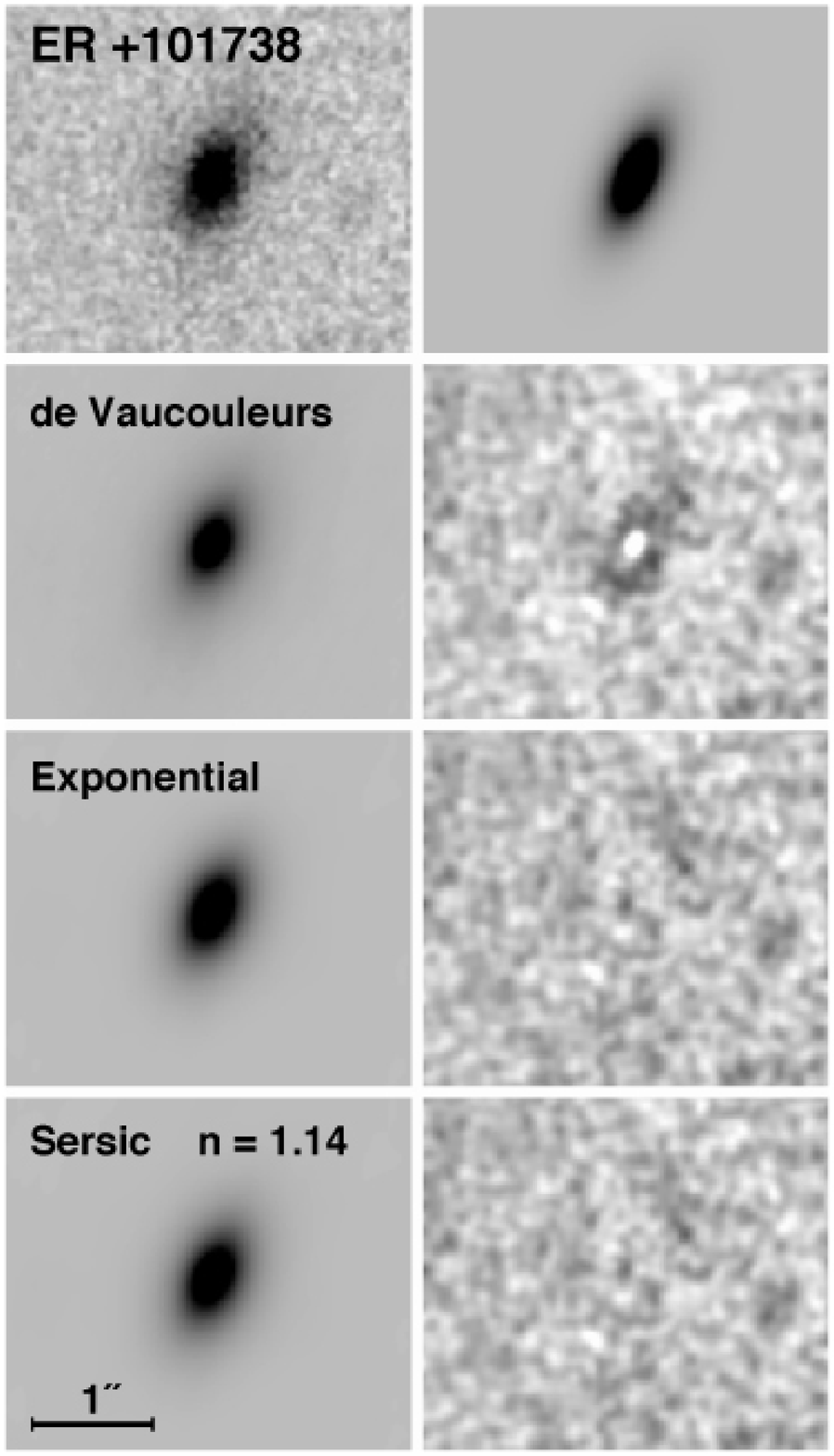}
\caption{Two-dimensional models of ER\,074941.6+101736 and ER\,074940.7+101738. 
The upper left panel in each case shows the 
original AO image; directly below are de Vaucouleurs (``$r^{1/4}$-law''), exponential, and
S\'{e}rsic models, each convolved with the PSF. To the right of each of the models is shown 
the smoothed residual from 
the subtraction of the  model from the original image. The upper-right panel shows
the S\'{e}rsic model {\it without} convolution with the PSF, which gives the best
realization of the gross morphology of the galaxy.}\label{gmod1736-8}
\end{figure}

\begin{figure}[!tb]
\epsscale{0.9}
\plottwo{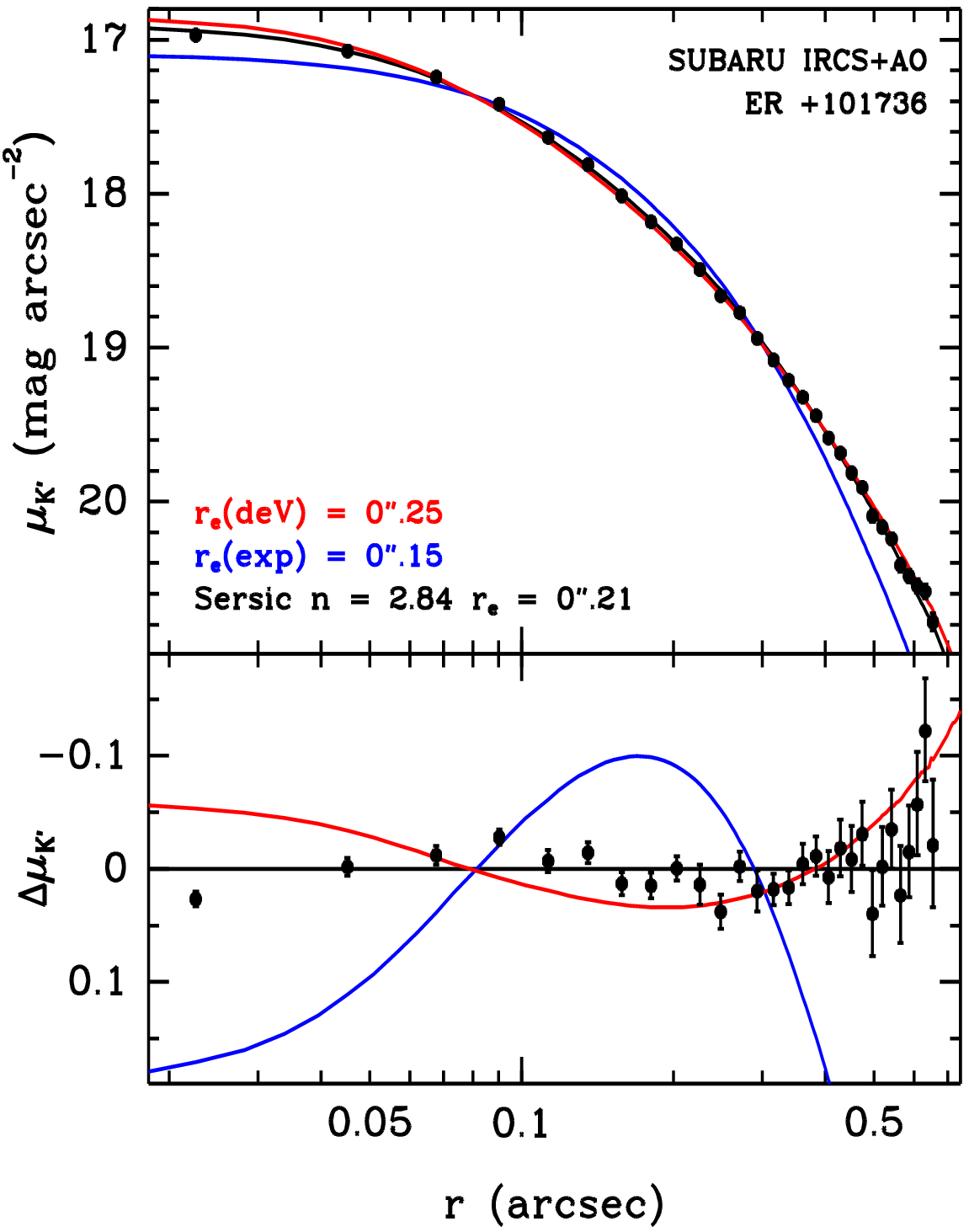}{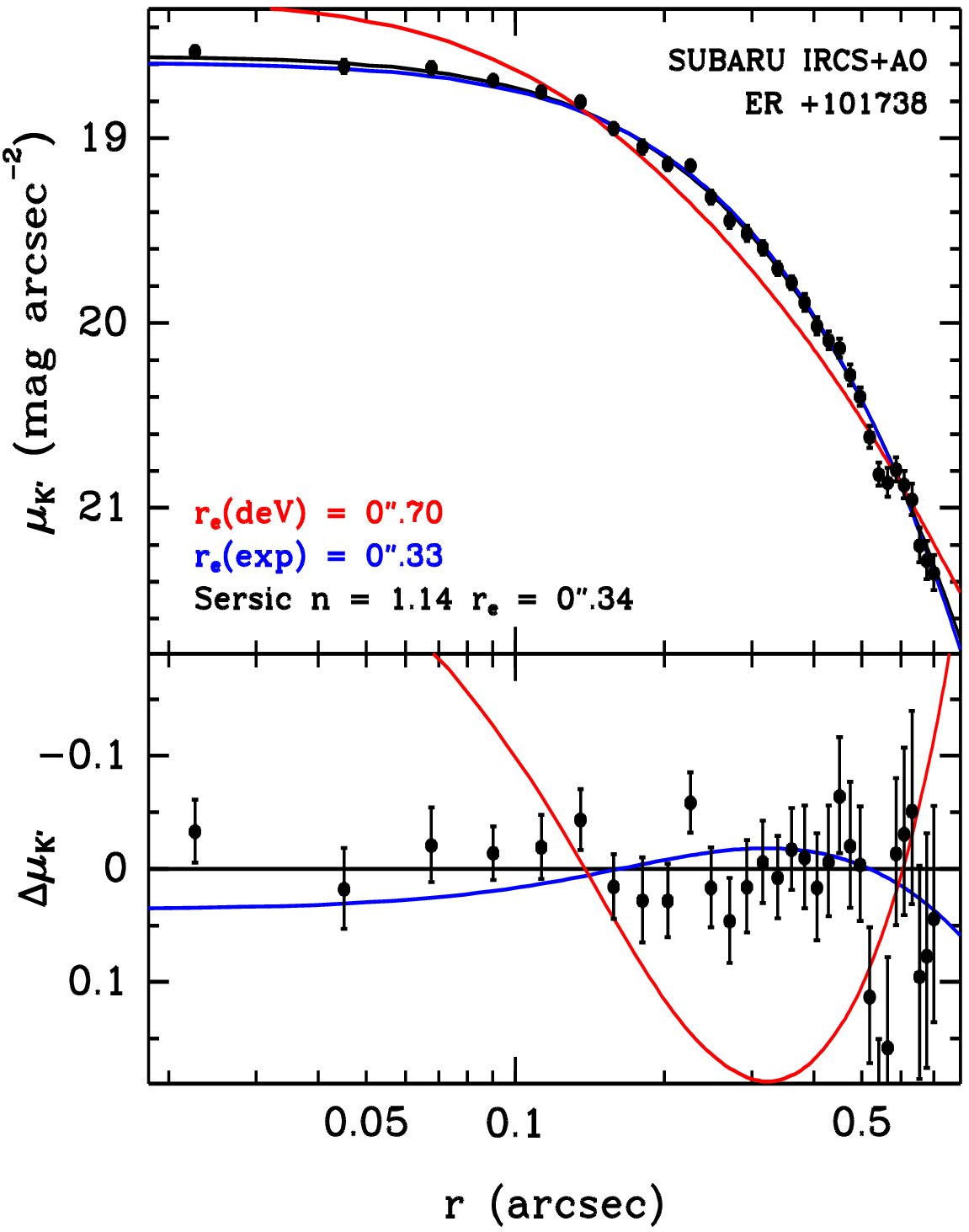}
\caption{PSF-convolved radial-surface-brightness profiles for ER\,074941.6+101736
and ER\,074940.7+101738. For each galaxy, the top panel shows 
the model profiles, coded as indicated by color, and with effective radii given for each of
the models. The bottom panel shows the deviations of the data points and the other
two models from the best S\'{e}rsic model.}\label{gsurf1736-8}
\end{figure}

This is the brightest of the extremely red galaxies in the AO field and the only one of these
four galaxies for which we have
a spectroscopic redshift. The two-dimensional model fits are shown on the left side of
Fig.~\ref{gmod1736-8}, and the corresponding radial-surface-brightness profiles are 
shown on the left side of Fig.~\ref{gsurf1736-8}. 

The exponential model is clearly a very poor fit, leaving large residuals at nearly all radii. 
The de Vaucouleurs model fits quite well overall, nearly as well as the best S\'{e}rsic model,
but it slightly oversubtracts the center. This behavior is consistent with a break in the
profile towards a flatter slope near the center. Examination of the lower part of
Fig.~\ref{gsurf1736-8} (left panel) indicates that one could tweak the de Vaucouleurs profile to obtain
an excellent fit to the observations beyond a semi-major axis of 0\farcs1 (840 pc), with
a flatter core at smaller radii. The behavior is similar to that of the most luminous ellipticals in the
local universe, most of which have ``core'' profiles (in the terminology of the Nuker
group: \citealt{fab97}; see also \citealt{kor99}). With only passive evolution, this galaxy would end up
slightly brighter than $L^*$ at the present epoch. An effective radius $r_e=0\farcs21$--0\farcs25
corresponds to 1.8--2.1 kpc, compared with a range of 1.7--4.0 kpc and a median of 2.9 kpc
for 4 local ellipticals with core profiles and similar luminosities in the sample of \citet{fab97}.

\subsection{ER\,074940.7+101738}
In this case, as is shown on the right sides of  Fig.~\ref{gmod1736-8} and Fig.~\ref{gsurf1736-8},
it is the de Vaucouleurs model that is clearly unacceptable. As the S\'{e}rsic
index of $n=1.14$ indicates, and as the surface-brightness profiles show, the profile is 
extremely close to a pure exponential at all radii.  This galaxy is likely to be an essentially
pure disk, with little or no bulge component. As is evident from the inset for this galaxy in
Fig.~\ref{ao}, the galaxy shows no well defined nucleus and weak central concentration.

\subsection{ER\,074941.0+101733}
\begin{figure}[!t]
\epsscale{0.9}
\plottwo{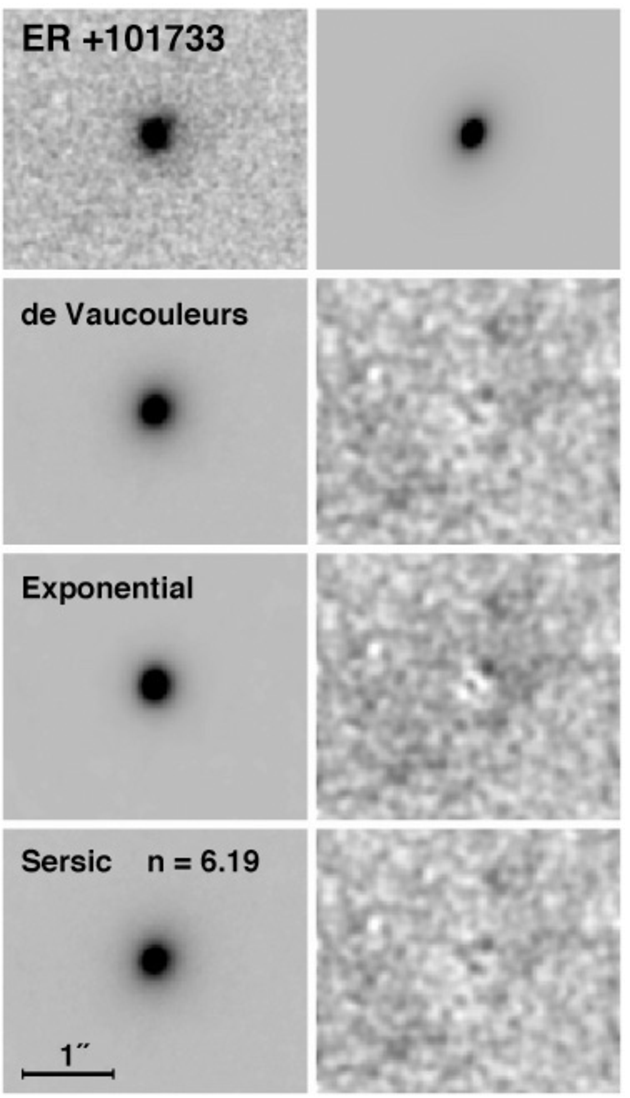}{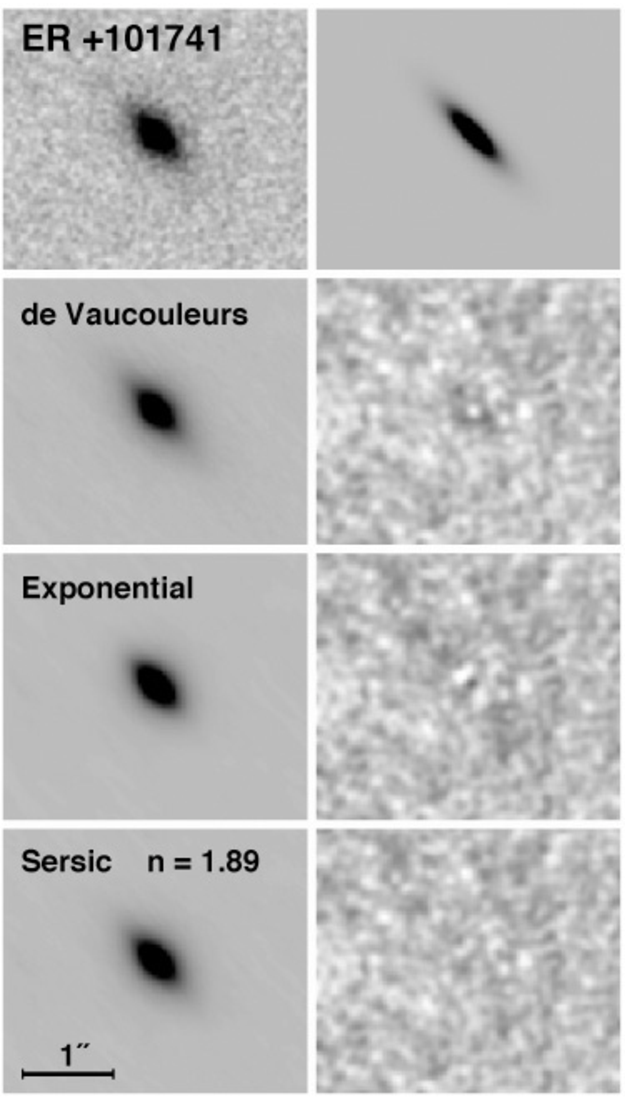}
\caption{Same as Fig.~\ref{gmod1736-8}, but for ER\,074941.0+101733 and
ER\,074941.4+101741.}\label{gmod1733-41}
\end{figure}

\begin{figure}[!tb]
\epsscale{0.9}
\plottwo{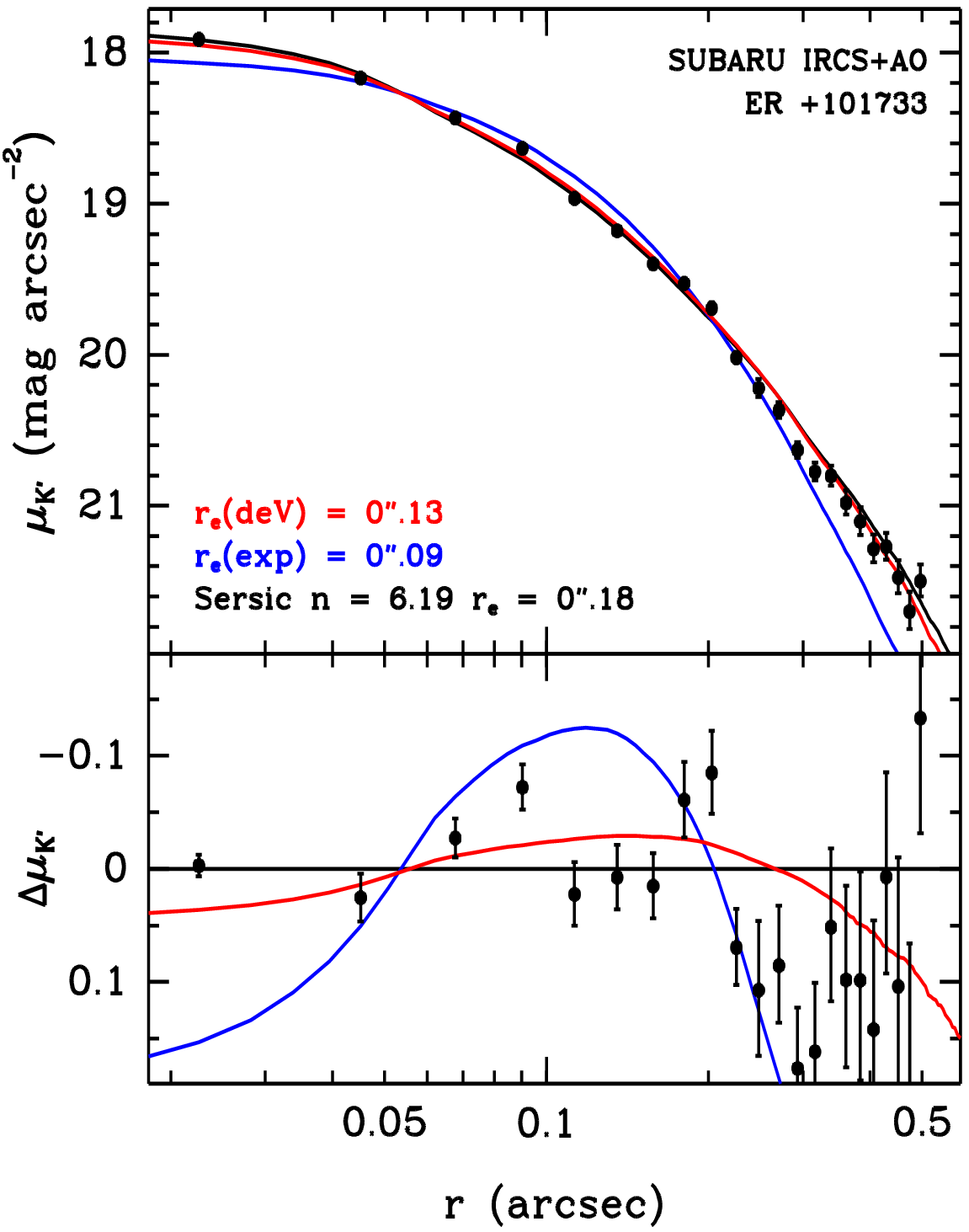}{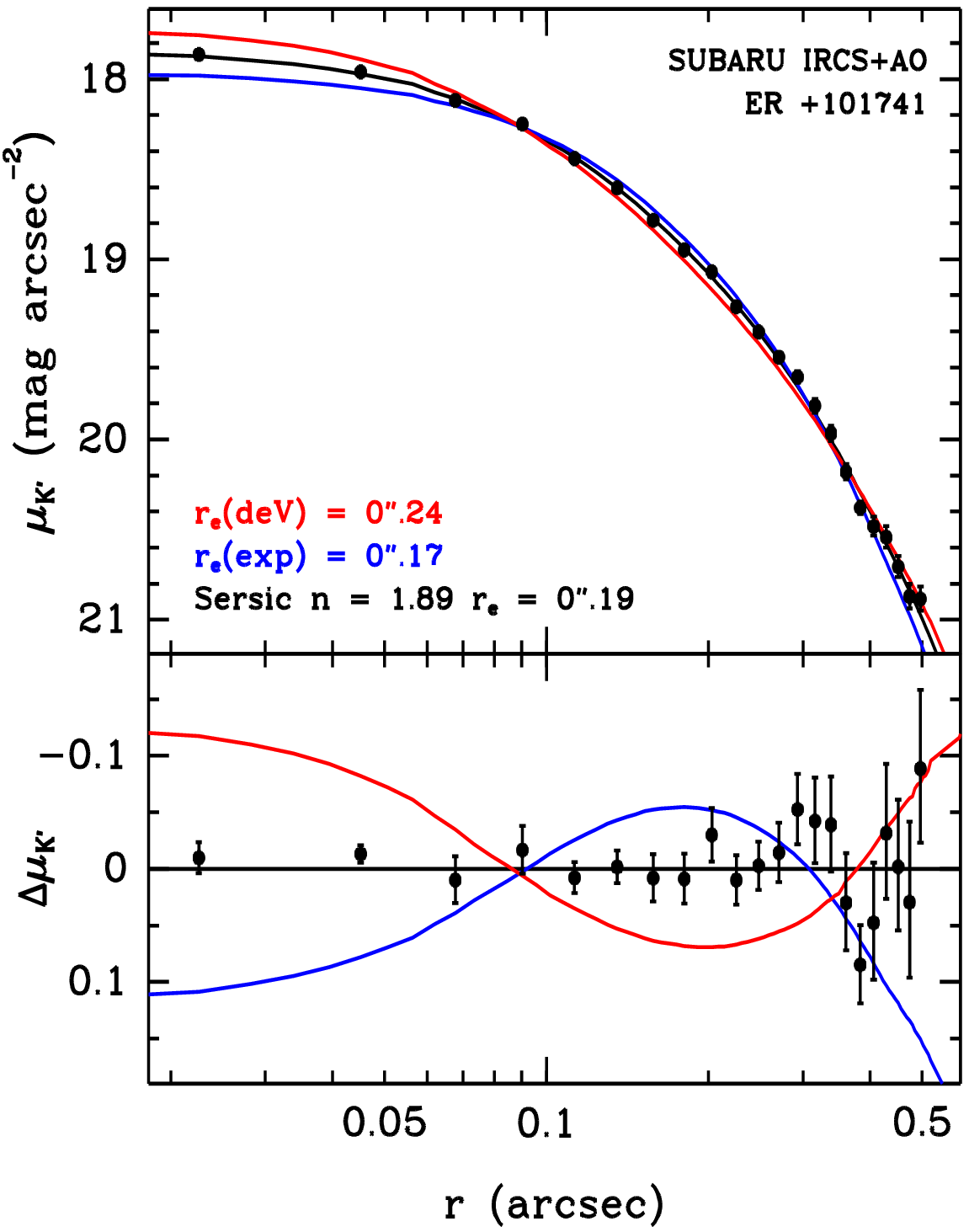}
\caption{Same as Fig.~\ref{gsurf1736-8}, but for ER\,074941.0+101733 and
ER\,074941.4+101741.}\label{gsurf1733-41}
\end{figure}

While the panels on the left sides of Fig.~\ref{gmod1733-41} and Fig.~\ref{gsurf1733-41} 
clearly show that the exponential 
fit for this galaxy is poor, there is little to choose from between
the de Vaucouleurs and the S\'{e}rsic profiles, even though the S\'{e}rsic profile with index
$n=6.2$ is formally a slightly better fit. The galaxy appears to be a small elliptical, with an
effective radius in the range of 1.0--1.5 kpc and a luminosity about a magnitude fainter than
$L^*$.

\subsection{ER\,074941.4+101741}
For this galaxy, Fig.~\ref{gmod1733-41} and Fig.~\ref{gsurf1733-41} show that the S\'{e}rsic 
model is much superior to either a pure de Vaucouleurs model
or a pure exponential model.  Nevertheless, the galaxy is highly flattened and clearly has
a dominant disk.  A composite model, incorporating an $r^{1/4}$-law bulge and an
exponential disk, fits slightly better than the best S\'{e}rsic model and puts 27\% of the
light in the bulge and 73\% in the disk.  The consistency of the SED with that of an old stellar population 
(see Fig.~\ref{sed}) indicates
that both the disk and bulge populations in this galaxy are composed of old stars.
All of these characteristics suggest a classification as an S0 galaxy much like
those we find at the present epoch.

\subsection{The Kormendy Relation}
Our high-resolution imaging of all of these galaxies shows them to be
quite regular and symmetric, and their basic morphological parameters appear to be within 
the range of those of local galaxies today. We can further ask whether the two
elliptical galaxies fall close to the mean
surface-brightness---effective radius (Kormendy) relation for local ellipticals.

For the bright elliptical galaxy ER\,074941.6+101736, we use the
de Vaucouleurs $r_e=0\farcs25$ and determine an average $K'$ surface brightness 
interior to $r_e$ of
$\langle\mu\rangle_e> = 17.55$ mag arcsec$^{-2}$.  This value must be corrected for
the difference in bandpass (observed $K'$ vs. rest-frame $V$), cosmological surface-brightness
dimming, and passive evolution.  We assume that the stellar population of the galaxy at the 
observed redshift can be modelled by a 3.5-Gyr \citet{bru03} solar-metallicity instantaneous burst.
Under passive evolution alone, this will correspond to an 12.35-Gyr solar-metallicity model at
the present epoch (we use the closest \citealt{bru03} model, at 12.25 Gyr). We normalize the
redshifted 3.5-Gyr model to give $K'=17.55$ through a synthetic $K'$ filter bandpass and
also normalize the rest-frame 12.25-Gyr model by the same factor (the redshifted and rest-frame
versions at each age are scaled to give the same bolometric flux).  We then
correct the 12.25-Gyr model for the $(1+z)^4$ surface-brightness dimming effect and
determine the $V$ magnitude, again from a synthetic $V$ filter bandpass, obtaining
$\langle\mu_0\rangle_e = 18.7$ mag arcsec$^{-2}$. The uncertainty in this determination
from the various assumptions and approximations we have made is at least 0.5 mag.

We compare this value with the determination of the local $V$-filter Kormendy relation given 
by \citet{laB03}, using the same cosmological parameters as we are assuming:
\begin{displaymath}
\langle\mu_0\rangle_e = 19.05 + 2.92 \log r_e\quad ,
\end{displaymath}
where $r_e$ is in kpc.  The uncertainty in this relation due to the combination of photometric
errors and the intrinsic scatter in this relation is $\sim0.5$ mag. For
ER\,074941.6+101736, $r_e=2.1$ kpc, giving $\langle\mu_0\rangle_e = 20.0$ mag arcsec$^{-2}$.
Thus, this galaxy, although seeming to have slightly higher surface brightness than expected, is
not far off the standard Kormendy relation, given the likely uncertainties.

We can carry out this same calculation for the fainter elliptical ER\,074941.0+10733. For this
galaxy, assuming a 2-Gyr age, we again obtain $\langle\mu_0\rangle_e = 18.8\pm0.5$ 
mag arcsec$^{-2}$. The Kormendy relation for this $r_e$ gives $19.2\pm0.5$, so this galaxy
falls on the relation within the expected uncertainty.

\section{Summary and Discussion}
\begin{figure}[!t]
\epsscale{1.0}
\plotone{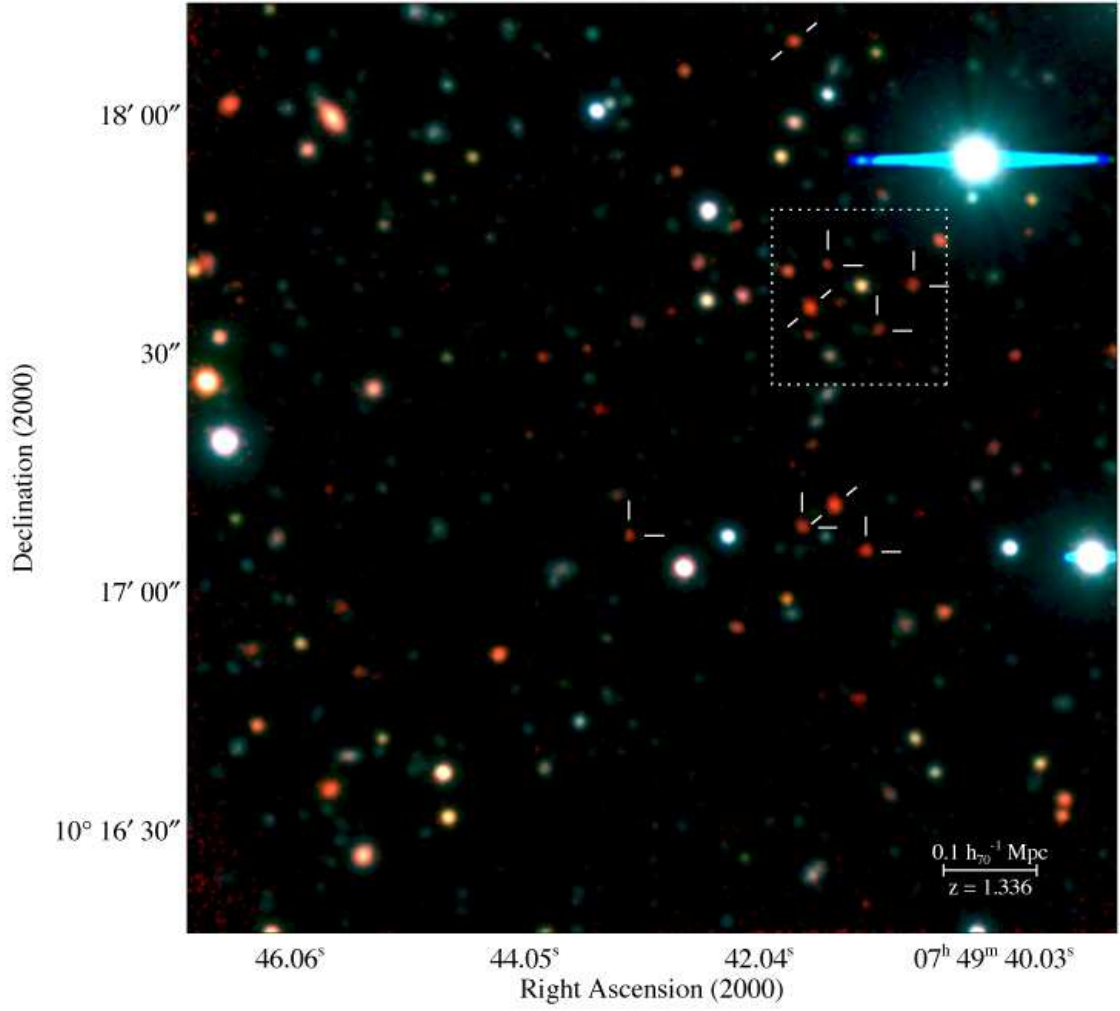}
\caption{The field of the ERO cluster.  The color table codes the $R$ image as blue, the $I$ 
image as green, and the $K'$ image as red. The images have been smoothed to a common
PSF.  The dotted white square indicates the AO field, 
and the saturated star to the upper right of the square is the AO guide star.
All marked objects have $R-K'\gtrsim6$; their SEDs shown in Fig.~\ref{sed}. There are clearly
other quite red objects in the field. The three objects indicated
by diagonal lines have spectroscopic redshifts (see Fig.~\ref{spect}). The angular size of the
scale bar is 11\farcs9.}\label{rik}
\end{figure}

\begin{figure}[!t]
\epsscale{0.55}
\plotone{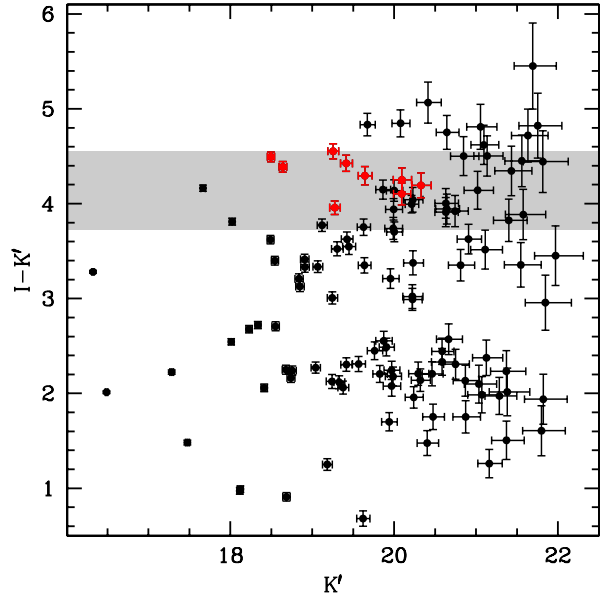}
\caption{$I\!-\!K'$ vs. $K'$ color-magnitude diagram of objects in the field shown in Fig.~\ref{rik}. 
The shaded band shows the region populated by ``red-sequence'' galaxies comprising pure old 
solar-metallicity stellar populations with ages ranging from 1 Gyr (bottom of band) to
3.5 Gyr (top of band) at a redshift of 1.34. The red points indicate the objects for which the
full range of photometry in all available bands gives SEDs consistent with 
very old stellar populations at $z\sim1.34$. The object in the middle of the band at $K'=17.7$
appears to have an old stellar population, but at a somewhat lower redshift. The objects
falling above the band at $K'\sim20$ tend to have SEDs characteristic of reddened
starbursts, as do some of the objects falling in the band.}\label{cmdiag}
\end{figure}

We have been loosely referring to IDSCL\,J0749+1018 as a cluster, although we admittedly 
have neither an extensive luminosity function nor X-ray data to back up this designation.
Until it is possible to obtain X-ray imaging, we can appeal to a comparison with 
ClG\,J0848+4453, at $z=1.27$  \citep{sta97}. This grouping of galaxies is confirmed as a
cluster by the presence of X-ray emission. This cluster has 6 galaxies with $K\gtrsim19.5$
and $R\!-\!K\gtrsim6$; these are galaxies which, through passive evolution alone, would be
expected to end up as $L^*$ galaxies at the present epoch.  As it happens, 
IDSCL\,J0749+1018 also has 6 such galaxies, so we tentatively identify it as a cluster
like ClG\,J0848+4453, but at a slightly higher redshift. As Fig.~\ref{rik} shows, there is
possible evidence for substructure: 7 of the 9 galaxies with $R\!-\!K\gtrsim6$ fall into two tight
groups of 4 and 3 galaxies, with projected dimensions enclosed within circles of 
15\arcsec\ (126 kpc) and 10\arcsec\ (84 kpc), respectively. Whether IDSCL\,J0749+1018
is truly a virialized cluster or not, it certainly seems to be the case that most of the galaxies
are in locally dense environments.

More importantly, we have obtained among the 
highest-resolution images at restframe $\sim0.9$ $\mu$m of a group of galaxies with already
old stellar populations when the universe was only a third of its present age. Significantly, 
even within this small sample of 4 galaxies, we find two galaxies that morphologically
appear to be standard ellipticals, one S0, and one that seems to be a pure disk. This 
variety reinforces the conclusions of \citet{yan04} that even EROs 
that are overwhelmingly dominated by old stellar populations show a variety of
morphologies. In particular, the presence of an essentially bulgeless disk comprising 
only old stars re-emphasizes 
that some disks in the early universe must have formed extraordinarily rapidly and
with high rates of star formation (cf.~\citealt{sto04}).

The other main conclusion of this work is that, unlike typical field galaxies at high redshift,
and with the exception of the apparent pure disk of old stars, the galaxies for which we 
have AO imaging look boringly similar to those we see around us today. It is quite
likely that these galaxies, all within a projected distance of 125 kpc of each other, will
suffer some structural changes from mergers and/or harassment (depending on their
velocity dispersion).  Nevertheless,  they are all individually well relaxed, symmetric 
stellar systems with properties that, through passive evolution alone, would make them
indistinguishable from a sample drawn from the present-day galaxy population. This
result tends to confirm the common assumption that processes of galaxy formation and
evolution are accelerated in denser environments.

\acknowledgments
This research has been partially supported by NSF grant AST03-07335. 
We are grateful to the other investigators in IfA-DS consortium: John Tonry,
Ken Chambers, Richard Wainscoat, Nick Kaiser, Pat Henry, Herv\'{e}
Aussel, Eduardo Mart\'{i}n, and Eugene Magnier; and to Brian Barris, who, 
with John Tonry, carried out almost all of the actual reduction of the survey images.
The authors recognize the very significant
cultural role that the summit of Mauna Kea has within the indigenous
Hawaiian community and are grateful to have had the opportunity to
conduct observations from it.

\begin{deluxetable}{lcccccc}
\tablewidth{0pt}
\tabletypesize{\footnotesize}
\tablecaption{Properties of Extremely Red Galaxies}
\tablehead{
\colhead{Galaxy} & \colhead{$R$} & \colhead{$I$} & \colhead{$Z$}
& \colhead{$J$} & \colhead{$K'$} & \colhead {Redshift} 
}

\startdata
ER\,074940.7+101738 & 25.6 +0.2,$-0.2$ & 23.81 $\pm0.05$ & 22.61 $\pm0.04$ & 21.5 $\pm0.2$ & 19.46  $\pm0.07$ & \nodata \\
ER\,074941.0+101733 & 26.5 +0.7,$-0.4$ & 24.23 $\pm0.07$ & 23.11 $\pm0.07$ & 21.5 $\pm0.2$ & 19.92  $\pm0.10$ & \nodata \\
ER\,074941.1+101705 & 26.4 +0.6,$-0.4$ & 23.93 $\pm0.06$ & 22.69 $\pm0.04$ & 21.2 $\pm0.2$ & 19.18  $\pm0.05$ & \nodata \\
ER\,074941.4+101711 & 24.9 +0.1,$-0.1$ & 23.07 $\pm0.03$ & 21.92 $\pm0.02$ & 20.4 $\pm0.1$ & 18.53  $\pm0.03$ & 1.338 \\
ER\,074941.4+101741 & 26.0 +0.4,$-0.3$ & 24.39 $\pm0.09$ & 23.06 $\pm0.06$ & 21.9 $\pm0.4$ & 20.04  $\pm0.11$ & \nodata \\
ER\,074941.6+101736 & 24.9 +0.1,$-0.1$ & 23.04 $\pm0.02$ & 21.96 $\pm0.02$ & 20.4 $\pm0.1$ & 18.64  $\pm0.03$ & 1.335 \\
ER\,074941.6+101708 & 25.3 +0.2,$-0.2$ & 23.83 $\pm0.05$ & 22.78 $\pm0.05$ & 21.4 $\pm0.2$ & 19.34  $\pm0.06$ & \nodata \\
ER\,074941.7+101809 & 24.8 +0.1,$-0.1$ & 23.09 $\pm0.03$ & 21.93 $\pm0.02$ & \nodata & 19.00 $\pm0.04$ & 1.336 \\
ER\,074943.1+101707 & 27.2 +2.6,$-0.7$ & 24.28 $\pm0.08$ & 23.29 $\pm0.08$ & 22.2 $\pm0.4$ & 19.84  $\pm0.09$& \nodata \\
\enddata
\label{galprop}
\end{deluxetable}



\end{document}